\documentclass[conference]{IEEEtran}
\IEEEoverridecommandlockouts
\usepackage{cite}
\usepackage{amsmath,amssymb,amsfonts}
\usepackage{algorithmic}
\usepackage{graphicx}
\usepackage{textcomp}
\usepackage{xcolor}
\usepackage[ruled,vlined,linesnumbered]{algorithm2e}
\usepackage{booktabs}
\usepackage{bm}
\usepackage{epstopdf}
\usepackage{enumitem}
\usepackage{glossaries}
\usepackage[colorlinks,linkcolor=red,anchorcolor=green,citecolor=blue]{hyperref}
\usepackage[switch]{lineno}
\usepackage{lscape}
\usepackage{url}
\usepackage{multirow}
\usepackage{makecell}
\usepackage{arydshln} 
\usepackage[tight,footnotesize]{subfigure}

\def\BibTeX{{\rm B\kern-.05em{\sc i\kern-.025em b}\kern-.08em
    T\kern-.1667em\lower.7ex\hbox{E}\kern-.125emX}}
\begin{document}

\title{Multiscale Spatio-Temporal Enhanced  Short-term Load Forecasting of Electric Vehicle Charging Stations\\
}

\author{\IEEEauthorblockN{1\textsuperscript{st} Zongbao Zhang}
\IEEEauthorblockA{\textit{Shenzhen Power Supply Bureau} \\
\textit{Co., Ltd.}\\
Shenzhen, China \\
zbzhangwhu@163.com}

\and
\IEEEauthorblockN{2\textsuperscript{nd} Jiao Hao}
\IEEEauthorblockA{\textit{Shenzhen Power Supply Bureau} \\
\textit{Co., Ltd.}\\
Shenzhen, China \\
haojiao.sjtu@163.com}

\and
\IEEEauthorblockN{3\textsuperscript{rd} Wenmeng Zhao}
\IEEEauthorblockA{\textit{Electric Power Research Institute,} \\
\textit{CSG}\\
Guangzhou, China \\
zhaowm@csg.cn}

\and
\IEEEauthorblockN{4\textsuperscript{th} Yan Liu}
\IEEEauthorblockA{\textit{Shenzhen Power Supply Bureau} \\
\textit{Co., Ltd.}\\
Shenzhen, China \\
liuyan0829@sz.csg.cn}

\and
\IEEEauthorblockN{5\textsuperscript{th} Yaohui Huang\IEEEauthorrefmark{1}}
\IEEEauthorblockA{\textit{Tsinghua-Berkeley Shenzhen Institute} \\
\textit{Tsinghua University}\\
Shenzhen, China \\
yhhuang5212@gmail.com}

\and
\IEEEauthorblockN{6\textsuperscript{th} Xinhang Luo}
\IEEEauthorblockA{\textit{Tsinghua-Berkeley Shenzhen Institute} \\
\textit{Tsinghua University}\\
Shenzhen, China \\
luoxh23@mails.tsinghua.edu.cn}

\thanks{(Corresponding author: Yaohui Huang, e-mail: yhhuang5212@gmail.com)

This paper was supported by the Science and Technology Project of Shenzhen Power Supply Corporation, grant number: 
SZKJXM20220036/09000020220301030901283.
}
}

\maketitle

\begin{abstract}

The rapid expansion of electric vehicles (EVs) has rendered the load forecasting of electric vehicle charging stations (EVCS) increasingly critical.
The primary challenge in achieving precise load forecasting for EVCS lies in accounting for the nonlinear of charging behaviors, the spatial interactions among different stations, and the intricate temporal variations in usage patterns.
To address these challenges, we propose a Multiscale Spatio-Temporal Enhanced Model (MSTEM) for effective load forecasting at EVCS. MSTEM incorporates a multiscale graph neural network to discern hierarchical nonlinear temporal dependencies across various time scales. Besides, it also integrates a recurrent learning component and a residual fusion mechanism, enhancing its capability to accurately capture spatial and temporal variations in charging patterns.
The effectiveness of the proposed MSTEM has been validated through comparative analysis with six baseline models using three evaluation metrics. The case studies utilize real-world datasets for both fast and slow charging loads at EVCS in Perth, UK.
The experimental results demonstrate the superiority of MSTEM in short-term continuous load forecasting for EVCS.

\end{abstract}


\section{Introduction}

The rapid development of electric vehicles (EVs) heralds a pivotal shift in the automotive sector \cite{richter2022circular}. Recently, the number of EVs has experienced a significant upsurge. By 2021, the global cumulative count of EVs reached approximately 16.5 million, a significant leap from the 5.5 million recorded in 2018 \cite{lin2023economic}. This burgeoning growth has led to a rapid increase in the demand for efficient and reliable electric vehicle charging stations (EVCS). Efficient management plays a key role in these stations, particularly in regulating the electric power supply. 
Accurate short-term forecasting of electrical loads is critical for maintaining power grid stability and enhancing the operational efficiency of EVCS.
However, forecasting the electric load for EVCS is a complex task. 
It involves numerous intricate factors, such as consumer usage patterns, time-of-day variations, options for fast and slow charging, and the volatility in the energy market \cite{he2022optimal}. 
These factors interact to create a dynamic and intricate demand scenario. 
Therefore, developing effective predictive models is essential to making accurate predictions and managing the EV charging demands.

There are various data-driven models that have been employed to forecast the electric loading at EVCS.
The statistical approaches, including Autoregressive Moving Average (ARMA) and Autoregressive Integrated Moving Average (ARIMA) models and their variations, have been widely used in charging load forecasting \cite{zhu2022review}. 
Lu et al. \cite{lu2021ultra} employ ARIMA in ultra-short-term demand response prediction.
Ren et al. \cite{ren2022hybrid} propose a hybrid network based on seasonal-ARIMA (SARIMA) to capture the periodic features of electric loads.
These models primarily focus on analyzing temporal charging patterns but often neglect the complex local correlations between charging stations within EVCS networks, which is difficult to accurately predict the charging patterns of EV charging demand.
As time series analysis increasingly integrates machine learning techniques, traditional models such as Multilayer Perceptron Network (MLP), Support Vector Machines (SVM), and Tree-based algorithms have shown significant potential in identifying EVs charging patterns.
Zhang et al. \cite{zhang2023hybrid} utilize the Gradient Boosting Decision Tree (GBDT) to enhance the predictive performance.
Khan et al. \cite{khan2023comparison} verify the effectiveness of traditional machine learning models in load forecasting for EVs.
Despite their improved reliability and accuracy, these models frequently struggle to capture the nonlinear temporal dynamics of charging demand, typically requiring extensive feature engineering for improved prediction. \cite{huang2023explainable}.
Deep learning is a promising model for extracting complex and nonlinear temporal patterns from electric loads, which includes Recurrent Neural Networks (RNN), Graph Neural Networks (GCNs), and attention-based models \cite{benidis2022deep}.
Wang et al. \cite{wang2023multivariate} utilized a graph attention mechanism to effectively learn the energy load patterns across multiple regions. 
Shi et al. \cite{journal/tsg2023/Shi} employed GCNs to capture the spatiotemporal relationships among charging stations. 
However, the above models fail to account for the dynamic interconnections between charging stations and the variations in temporal details across different scales.

The major challenges in short-term load forecasting for EVCS can be summarized into three aspects: 
(1) \textit{Complex Loading Patterns}. 
EVCS charging loads display significant nonlinear fluctuations influenced by various factors, such as weather conditions, commuting patterns, and social events. Therefore, it is a significant challenge to accurately capture these complex temporal patterns.
(2) \textit{Spatial Dynamics of EVCS}. 
The load trends of adjacent EVCS often exhibit similarities, highlighting the influence of spatial factors on charging behaviors. The location of each significantly affects its usage patterns. These spatial dynamics are crucial for predicting charging station load trends.
(3) \textit{Hierarchical Temporal Patterns}. 
The electric load exhibits distinct patterns across different temporal scales \cite{spiliotis2020cross}. These patterns range from short-term fluctuations to long-term trends, presenting a challenge for prediction models to effectively capture and interpret these diverse temporal interactions.

To tackle the above challenges in short-term load forecasting for EVCS, we propose a novel Multiscale Spatio-temporal Enhanced Model (MSTEM).
For the first challenge, MSTEM incorporates an LSTM unit to capture the nonlinear temporal dependencies from the historical load patterns.
To tackle the second challenge, MSTEM incorporates GCNs to effectively consider the local correlations among charging stations.
For the third challenge, MSTEM implements a multi-scale fusion strategy that captures the dynamics of the graph structure across various temporal scales. This approach enables the model to accurately discern complex load patterns at different time scales.
The main contributions of this paper are summarized as follows:
\begin{itemize}
	\item A novel neural network architecture, termed MSTEM, has been proposed for load forecasting at EVCS. MSTEM employs a multi-scale graph neural network with recurrent learning components to effectively capture linear trends and nonlinear temporal representations.
	
	\item The effectiveness of the proposed model is validated using a real-world dataset encompassing both fast and slow charging scenarios for EVCS.  
 
    \item Comparative experimental results indicate that MSTEM surpasses established baseline models, including Dlinear, MLP, and variants of RNN, in terms of forecasting accuracy over multiple time steps
\end{itemize}

The rest of the article is organized as follows: Section \ref{section:method} provides the mathematical formulation of the proposed MSTEM, followed by Section \ref{section:experiment} presents the case studies and results; Section \ref{section:conclusion} concludes the article.

\section{The framework of Multiscale Spatio-temporal Enhanced Model (MSTEM)}
\label{section:method}

\subsection{Problem definition}

The load forecasting at EVCS is essentially a time series forecasting problem, which aims to predict future load values based on historical observations.
In this study, the objective of charging load forecasting is to utilize the observed history of the previous $\tau$ time steps to predict the values of the target time series for the next $\alpha$ time steps. This multi-step continuous forecasting problem can be expressed as follows:
\begin{equation}
    \begin{aligned}
        [x_t, x_{t+1}, \ldots, x_{t+\tau-1}] \xrightarrow[\theta]{\mathbb{F}(\cdot)} [x_{t+\tau},  \ldots, x_{t+\tau+\alpha-1}],
    \end{aligned}
\end{equation}
where $x_i \in \mathbb{R}^{N}$ represents the charging load value at the $i$-th time step. 
$N$ is the number of target EVCS.
$X_t = [x_t, x_{t+1}, \ldots, x_{t+\tau-1}] \in \mathbb{R}^{\tau \times N}$ denotes the model inputs, and the subsequent sequence $Y_t = [x_{t+\tau},  \ldots, x_{t+\tau+\alpha-1}] \in \mathbb{R}^{\alpha \times N}$ is the predictive target. $\mathbb{F}(\cdot)$ indicates the predictive model with learnable parameter $\theta$.

\subsection{The Overview of MSTEM}

\begin{figure*}[t!]
	\centering
	\includegraphics[width=0.95\linewidth]{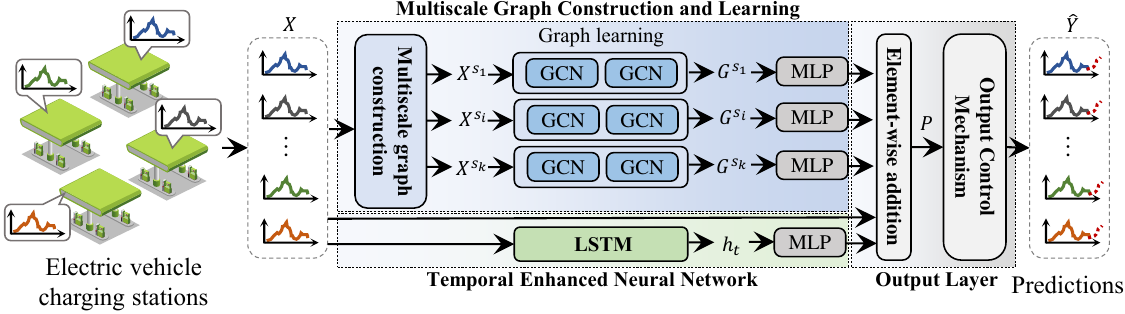}
	\vspace{-10pt}
	\caption{The overall structure of the multiscale spatio-temporal enhanced model.}
	\vspace{-15pt}
	\label{fig:method}
\end{figure*}

The proposed MSTEM, depicted in Fig. \ref{fig:method}, consists of three sequential modules: Multiscale Graph Construction and Learning (MGCL), Temporal Enhanced Neural Network (TENN), and an output layer.
The MGCL module focuses on capturing the spatio-temporal dependencies among different charging stations, utilizing a graph-based approach to model the intricate interactions and patterns in data.
The TENN module combines the LSTM component with linear representations, which are derived from raw input data. This integration facilitates a comprehensive understanding of temporal patterns.
The output layer incorporates output control mechanisms to refine and adjust the predicted outcomes based on the processed data and learned patterns.

\subsection{Multiscale Graph Construction and Learning (MGCL)}
For an input charging load record $X \in \mathbb{R}^{\tau \times N}$ from multiple charging stations, 
the multiscale load patterns can be defined as:
\begin{equation}
	\begin{aligned}
		X^{s_i} = \text{ScaleTemporal}(X, s_i)~~~~~ \forall s_i \in S,
	\end{aligned}
\end{equation}
where $X^{s_i}$ denotes the scaled input time series.
$s_i$ is an element within the set of scale lengths, denoted as $S=\{s_1, s_2, ..., s_k\}$.
The function \text{ScaleTemporal}$(\cdot)$ samples the input data $X$ at intervals of $s_i$.
For instance, if $s_i=2$, it selects every second time step from $X$.
Then, MSTEM performs graph convolution on each scaled input. 
The time series of charging load can be seen as a completed graph, let $A^{s_i}$ represents the adjacency matrix and $D^{s_i}$ denotes the corresponding degree matrix. The graph convolution for scale $s_i$ is:
\begin{equation}
	\begin{aligned}
		H^{s_i} = \sigma(\text{BN}(\tilde{D}^{-\frac{1}{2}} \tilde{A} \tilde{D}^{-\frac{1}{2}} X^{s_i} W^{s_i}_1)),
	\end{aligned}
\end{equation}
\begin{equation}
	\begin{aligned}
		R^{s_i} = \text{Dropout}(H^{s_i}_1; p),
	\end{aligned}
\end{equation}
\begin{equation}
	\begin{aligned}
		G^{s_i} = \tilde{D}^{-\frac{1}{2}} \tilde{A} \tilde{D}^{-\frac{1}{2}} R^{s_i} W^{s_i}_2,
	\end{aligned}
\end{equation}
where $H^{s_i}$ and $G^{s_i}$ is the hidden and final representation of graph learning at scale $s_i$. 
$\tilde{A} = A + I$ includes self-loops, $I$ is the unit matrix. $\tilde{D}$ is the degree matrix of $\tilde{A}$. $W^{s_i}$ is the weight matrix for scale $s_i$, and $\sigma(\cdot)$ represents the Rectified Linear Unit (ReLU) activation function. $p$ is the dropout rate. $\text{BN}(\cdot)$ represents the batch normalization layer.
After processing the scaled data, MSTEM integrates the features from each scale. 
The features from each scale are linearly transformed to a consistent shape, and then are combined through element-wise addition:
\begin{equation}
	\begin{aligned}
		G^{out} =\sum_{i=1}^{k} G^{s_i}_2 W_L^{s_i},
	\end{aligned}
\end{equation}
where $G^{out}$ denotes the combined feature vector, which integrates information from all scales.

\subsection{Temporal Enhanced Neural Network (TENN)}
The MGCL module excels in learning the spatial and temporal information from the charging load data at EVCS. However, certain limitations need to be addressed.
(1) The MGCL primarily focuses on capturing spatial dependencies and temporal patterns at different scales but may not effectively model long-term temporal dependencies.
(2) The learning process within the MGCL module may result in information loss, particularly when handling complex nonlinear representations.
To solve these issues, the TENN module integrates LSTM units to enhance the model's ability to capture long-term temporal dependencies. LSTMs are well-suited for processing time-series data and can retain information over extended periods, which is essential for accurate EVCS load forecasting. 
This processing can be mathematically represented as:
\begin{align}
	f_t &= \sigma(W_f^{l} \cdot [h_{t-1}, x_t] + b_f^{l}), \\
	i_t &= \sigma(W_i^{l} \cdot [h_{t-1}, x_t] + b_i^{l}), \\
	\tilde{c}_t &= \tanh(W_C^{l} \cdot [h_{t-1}, x_t] + b_C^{l}), \\
	c_t &= f_t \odot c_{t-1} + i_t \odot \tilde{c}_t, \\
	o_t &= \sigma(W_o^{l} \cdot [h_{t-1}, x_t] + b_o^{l}), \\
	h_t &= o_t \odot \tanh(c_t),
\end{align}
where $f_t$, $i_t$, and $o_t$ are the output of the forget gate, input gate, and output gate at time step $t$, respectively.
$\tilde{c}_t$ and $c_t$ are the candidate, updated cell state, respectively.
$h_t$ is the updated hidden state.
$\odot$ denotes element-wise multiplication. $W^{l}$ and $b^{l}$ are the weights and biases for the respective gates and states.

Additionally, the proposed MSTEM integrates a residual fusion mechanism to preserve the integrity of the original time series signal. This mechanism introduces linear characteristics into the model and plays a crucial role in stabilizing the learning process. As a result, it significantly improves the robustness and reliability of the proposed MSTEM. The linear representations within this mechanism can be expressed as:
\begin{equation}
	\begin{aligned}
		L^{out} = \tilde{X}^{\eta} \oplus (XW^{lo}),
	\end{aligned}
\end{equation}
where $\tilde{X}^{\eta}$ denotes a historical segment of the charging load data, capturing the sequence from the past 
$\eta$ time steps, specifically including the data points $\{x_{t+\tau-\eta-1}, \dots, x_{t+\tau-1}\}$.  $W^{lo}$ is a learnable weighting matrix designed to modify the dimensions of $X$ to align with the desired output shape.  $\oplus$ represents element-wise addition.
 
\subsection{Output Layer} The output layer of the model combines the outputs from the MGCL and TENN modules to generate predictions, which is formulated as:
\begin{equation}
	\begin{aligned}
		P = (W^{out}_1 G^{out}) \oplus (W^{out}_2 h_t) \oplus L^{out},
	\end{aligned}
\end{equation}
where $P \in \mathbb{R}^{\alpha \times N}$ represents the charging load prediction of each station in the next $\alpha$ time steps. $W^{out}$ is utilized to linearly transform the outputs of the aforementioned modules. To align with real-world operation, an output control mechanism is integrated to modify the extremely low predictions to zero:
\begin{equation}
	\hat{Y} = \max(P_i - \theta, 0) ~~~  i \in \{t+\tau, \dots, t+\tau+\alpha-1\},
\end{equation}
where $\hat{Y}$ denotes the final predictions, and $\theta$ is a predefined threshold.
 
\section{Case study}
\label{section:experiment}

\subsection{Experimental Setting}

\subsubsection{Datasets} 
To evaluate the accuracy of our proposed model,
real-world electric vehicle charging load datasets from Perth, UK \footnote{\url{https://data.pkc.gov.uk/}} were analyzed in this study.
%and Boulder \footnote{\url{https://open-data.bouldercolorado.gov/datasets/}} and Palo Alto, USA \footnote{\url{https://data.cityofpaloalto.org/dataviews/241685/ELECT-VEHICCHARG-STATI-USAGE/}}.
This dataset includes the start and end times, and the total energy consumption for each charging event. 
To meet the requirements of load forecasting, the original dataset is transformed into an hourly resolution time series, representing the average charging load per hour.
This dataset was also utilized in prior research \cite{journal/access2023/11Mohammad}, although for different research objectives. 
The dataset was separated into two subsets according to the charging equipment type: a fast charging dataset and a slow charging dataset. The fast charging subset comprises equipment characterized by a power output of approximately 22 kW. In contrast, the slow charging subset consists of equipment with power outputs around 7 kW or lower. 
%Figure \ref{fig:dateset} displays the 10-day charging load curves sampled from each dataset.
%The basic statistics of these two subsets are presented in Table \ref{table:statistical}.
%
%\begin{table}[htbp]
%  \centering
%  \caption{The statistics of charging load datasets.}
%  \resizebox{1\linewidth}{!}{
%    \begin{tabular}{ccccccc}
%    \toprule
%    Dataset & \makecell[c]{Number of \\ data points} & \makecell[c]{Average \\ load \\ ($kW$)} & \makecell[c]{Median \\ load \\ ($kW$)} & \makecell[c]{Peak \\ load \\ ($kW$)} & \makecell[c]{None-zero charging \\ hour ratio \\ ($\%$)} & \makecell[c]{Standard \\ deviation \\ ($kW$)} \\
%    \midrule
%    Boulder & 22632 & 6.78  & 3.96  & 53.12  & 64.74  & 8.29  \\
%    Palo Alto & 3113  & 2.01  & 0.41  & 16.43  & 52.39  & 2.78  \\
%    Perth & 17520 & 13.95  & 7.35  & 124.91  & 95.36  & 16.30  \\
%    \bottomrule
%    \end{tabular}%
%    }
%  \label{table:statistical}%
%\end{table}%

% \begin{figure} [t!]
%     \centering
%     \includegraphics[width=0.85\linewidth]{./images/charging_stations.pdf}
%     \vspace{-10pt}
%     \caption{The visualization of 10-day charging load curves of three different charging stations.}
%     \vspace{-10pt}
%     \label{fig:dateset}
% \end{figure}

%, highlighting several forecasting challenges: (a) \textit{Non-Stationarity}, with substantial load value fluctuations including sudden, unpredictable surges in charging demand; (b) \textit{Data Sparsity}, characterized by intervals of little to no charging activity. These challenges constrain the accuracy of forecasting methods.

\subsubsection{Implementation Details}
The experimental setup was implemented on a server equipped with an NVIDIA A6000 GPU. The dataset was divided into three parts: the first 70\% of the historical load data is used for training, 10\% for validation, and the remaining 20\% for testing. We train the model using the Adam optimizer and the Huber loss function, setting the learning rate at 0.001. The training process involved a batch size of 32 and is conducted over 50 epochs. For the proposed MSTEM, the scale parameters are set to $\{1, 5\}$. The architecture of each graph learning component included two sequentially stacked graph convolutional modules. The dimensions for the hidden size and graph output layers are set at 16 and 4, respectively. The dropout rate $p$ is 0.1.

\subsubsection{Performance Metrics}

To evaluate the performance of the proposed model in charging load forecasting, three metrics are employed: mean square error (MSE), mean absolute error (MAE), and root mean square error (RMSE). 
Specifically, MSE emphasizes larger errors, MAE indicates the average magnitude of errors, and RMSE quantifies the standard deviation of the forecasting errors.
The formulations of the above three metrics are provided below:
\begin{equation}
    \begin{aligned}
        \text{MSE} = \frac{1}{N \cdot \Omega_{test}} \sum_{i=1}^{N} \sum_{j=1}^{\Omega_{test}} (\hat{Y}_{i,j} - Y_{i, j})^2,
    \end{aligned}
\end{equation}
\begin{equation}
    \begin{aligned}
        \text{MAE} = \frac{1}{N \cdot \Omega_{test}} \sum_{i=1}^{N} \sum_{j=1}^{\Omega_{test}} |\hat{Y}_{i,j} - Y_{i, j}|,
    \end{aligned}
\end{equation}
\begin{equation}
    \begin{aligned}
        \text{RMSE} = \sqrt{\frac{1}{N \cdot \Omega_{test}} \sum_{i=1}^{N} \sum_{j=1}^{\Omega_{test}} (\hat{Y}_{i,j} - Y_{i, j})^2},
    \end{aligned}
\end{equation}
where $\hat{Y}$ and $Y$ represent the predicted and actual values, respectively. $\Omega_{test}$ denotes the number of samples in the test set. Lower metric scores indicate greater model accuracy.

\subsection{Accuracy Comparison}

In this study, we assess the performance of our proposed model using datasets encompassing both fast and slow charging loads at EVCS. These datasets are analyzed over diverse time steps ($\alpha$): 6 hours, 12 hours, and 24 hours. The results of this analysis are presented in Table \ref{table:exp_fast} for fast charging and in Table \ref{table:exp_slow} for slow charging. 
The best result for each metric is emphasized in bold, while the second-highest is distinguished with an underline.
To evaluate the effectiveness of the proposed model, six baseline models are selected for comparative analysis against the proposed MSTEM model, as follows:
\begin{itemize}
	\item Moving average (\textbf{MA}) \cite{kaytez2020hybrid}: A classical statistical technique that computes the average of data points within a specified time window.
	\item Historical inertia (\textbf{HI}) \cite{conference/cikm2021/2965cui}: This model utilizes the inertia or momentum inherent in historical data to make predictions.
	\item Multilayer perception network (\textbf{MLP}) \cite{dudek2020multilayer}: A type of feedforward artificial neural network capable of modeling complex nonlinear relationships in charging load data.
	\item Gated recurrent unit (\textbf{GRU}) \cite{wang2018short}: A variant of the traditional RNN incorporates gating mechanisms to address the vanishing gradient problem.
	\item Long short-term memory (\textbf{LSTM}) \cite{8312088}: This model is distinguished by its specially designed memory cells based on RNN architecture. These cells enable the network to retain information over prolonged durations.
	\item Dlinear \cite{zeng2023transformers}: This model focuses on linear relationships in time series forecasting. It is particularly effective in scenarios where the data exhibits linear trends.
\end{itemize}

\begin{table}[t!]
	\centering
	\caption{Results of comparable experiments on fast charging dataset.}
	\resizebox{1\linewidth}{!}{
	    \begin{tabular}{ccccccccc}
			\toprule
			Step & Metrics & MA    & HI    & MLP   & GRU   & LSTM  & Dlinear & MSTEM \\
			\midrule
			\multirow{3}[2]{*}{6 h} & MSE   & 36.89  & 51.37  & \textbf{22.19}  & 24.15  & 23.63  & 22.49  & \underline{22.33}  \\
			& MAE   & 2.74  & 2.97  & \underline{2.18}  & 2.25  & 2.24  & 2.24  & \textbf{2.07}  \\
			& RMSE  & 6.07  & 7.17  & \textbf{4.71}  & 4.91  & 4.86  & 4.74  & \underline{4.73}  \\
			\midrule
			\multirow{3}[2]{*}{12 h} & MSE   & 42.25  & 53.02  & \underline{23.03}  & 24.68  & 24.11  & 23.20  & \textbf{22.95}  \\
			& MAE   & 3.16  & 3.06  & \underline{2.22}  & 2.31  & 2.22  & 2.27  & \textbf{2.15}  \\
			& RMSE  & 6.50  & 7.28  & \underline{4.80}  & 4.97  & 4.91  & 4.82  & \textbf{4.79}  \\
			\midrule
			\multirow{3}[2]{*}{24 h} & MSE   & 29.14  & 54.08  & \underline{23.41}  & 30.54  & 30.47  & 23.53  & \textbf{22.61}  \\
			& MAE   & 2.64  & 3.12  & \underline{2.27}  & 3.06  & 3.08  & 2.28  & \textbf{2.15}  \\
			& RMSE  & 5.40  & 7.35  & \underline{4.84}  & 5.53  & 5.52  & 4.85  & \textbf{4.75}  \\
			\bottomrule
		\end{tabular}%
	}
	\label{table:exp_fast}%
\end{table}%

For the results from the fast charging dataset in Table \ref{table:exp_fast}, the proposed MSTEM model demonstrates superior performance across various metrics and forecasting steps. Specifically, MSTEM achieves the lowest MSE, MAE, and RMSE scores in the 12 h and 24 h forecasting steps, with MSE scores of 22.95 and 22.61 and RMSE scores of 4.79 and 4.75, respectively. The MLP model also performs relatively higher accuracy in results, particularly in minimizing MSE, whereas the HI model generally exhibits lesser performance. Both the GRU and LSTM models demonstrate moderate effectiveness, which considers the limited electric load patterns and temporal dependencies. The Dlinear model, focusing on linear relationships, shows commendable results, particularly in shorter forecast steps. 
These results highlight the robust predictive accuracy and reliability of MSTEM in fast charging scenario. The consistent outperformance in MSE, MAE, and RMSE underscores its effectiveness in handling the complexities associated with fast charging load forecasting.

\begin{table}[t!]
	\centering
	\caption{Results of comparable experiments on slow charging dataset.}
	\resizebox{1\linewidth}{!}{
    \begin{tabular}{ccccccccc}
		\toprule
		Step & Metrics & MA    & HI    & MLP   & GRU   & LSTM  & Dlinear & MSTEM \\
		\midrule
		\multirow{3}[2]{*}{6 h} & MSE   & 2.64  & 3.43  & 1.54  & 1.61  & 1.61  & \textbf{1.52}  & \textbf{1.52}  \\
		& MAE   & 0.57  & 0.60  & 0.48  & 0.60  & 0.64  & \underline{0.50}  & \textbf{0.45}  \\
		& RMSE  & 1.63  & 1.85  & 1.24  & 1.27  & 1.27  & \textbf{1.23}  & \textbf{1.23}  \\
		\midrule
		\multirow{3}[2]{*}{12 h} & MSE   & 3.08  & 3.64  & 1.67  & 1.89  & 2.03  & \textbf{1.64}  & \underline{1.66}  \\
		& MAE   & 0.71  & 0.64  & \underline{0.51}  & 0.69  & 0.70  & 0.53  & \textbf{0.49}  \\
		& RMSE  & 1.75  & 1.91  & \textbf{1.29}  & 1.38  & 1.43  & \textbf{1.28}  & \underline{1.29}  \\
		\midrule
		\multirow{3}[2]{*}{24 h} & MSE   & 2.13  & 3.76  & \underline{1.72}  & 2.08  & 2.08  & \textbf{1.70}  & 1.77  \\
		& MAE   & 0.60  & 0.67  & \underline{0.52}  & 0.71  & 0.71  & 0.53  & \textbf{0.50}  \\
		& RMSE  & 1.46  & 1.94  & \underline{1.31}  & 1.44  & 1.44  & \textbf{1.30}  & 1.33  \\
		\bottomrule
	\end{tabular}%
	}
	\label{table:exp_slow}%
\end{table}%

Table \ref{table:exp_slow} presents the comparative analysis results from the slow charging dataset. 
In the 6 h forecasting step, Dlinear and MSTEM both achieve the lowest MSE score of 1.52. MSTEM obtains the best MAE at 0.45 and tying with Dlinear for the top RMSE score of 1.23.
In the 12 h and 24 h forecasting step, MSTEM maintains the lead in MAE with 0.49 and 0.50, respectively. 
Dlinear performs well in MSE and RMSE, which indicating its strength in shorter-term forecasts. 
These results highlights the effectiveness of Dlinear and MSTEM in slow charging predictions. In contrast, the HI and MA models show lower performance, and GRU and LSTM exhibit moderate results.
A possible reason is that these models have a potential mismatch with the distinct characteristics of slow charging data, which typically involves longer charging cycles and less variability compared to fast charging.

Overall, the experimental results of both fast and slow charging datasets demonstrates that the proposed MSTEM outperforms other models, especially in MAE and RMSE metrics. The superior performance highlights the predictive accuracy and reliability of the proposed MSTEM.

%\subsection{Visualization of Forecasting}

\section{Conclusions}
\label{section:conclusion}

This paper proposes the MSTEM to enhance short-term load forecasting at EVCS.
MSTEM effectively integrates a LSTM network to adeptly handle the complex temporal dynamics of EVCS electric loads, capturing nonlinear long-term dependencies.
It further leverages a residual fusion approach to supplement its analysis of linear characteristics.
Additionally, MSTEM accounts for the interrelationship of different charging stations and the hierarchical temporal patterns in charging loads, employing a multiscale graph neural network to discern spatial dynamics across various time scales.
The efficacy of MSTEM has been validated through datasets encompassing both fast and slow charging loads at EVCS in Perth, UK.
The experimental results demonstrate that the proposed MSTEM outperforms other models across multiple evaluation metrics.

% To consider the complex temporal patterns of the electric loads at EVCS, MSTEM incorporates LSTM to capture the nonlinear long-term dependencies and utilize the residual fusion mechanism to complement the linear characteristics. Besides, MSTEM considers the relationship among different charging stations and hierarchical temporal patterns of charging loads, employing the multiscale graph neural network to capture the spatial dynamics in different time scales. The case studies are conducted on the dataset from fast and slow charging loads from EVCS in Perth, UK. The experimental results demonstrate that the proposed MSTEM outperforms other models across multiple evaluation metrics.

% \section*{Acknowledgment}

\bibliography{evc.bib}

\begin{thebibliography}{10}
\providecommand{\url}[1]{#1}
\csname url@samestyle\endcsname
\providecommand{\newblock}{\relax}
\providecommand{\bibinfo}[2]{#2}
\providecommand{\BIBentrySTDinterwordspacing}{\spaceskip=0pt\relax}
\providecommand{\BIBentryALTinterwordstretchfactor}{4}
\providecommand{\BIBentryALTinterwordspacing}{\spaceskip=\fontdimen2\font plus
\BIBentryALTinterwordstretchfactor\fontdimen3\font minus
  \fontdimen4\font\relax}
\providecommand{\BIBforeignlanguage}[2]{{%
\expandafter\ifx\csname l@#1\endcsname\relax
\typeout{** WARNING: IEEEtran.bst: No hyphenation pattern has been}%
\typeout{** loaded for the language `#1'. Using the pattern for}%
\typeout{** the default language instead.}%
\else
\language=\csname l@#1\endcsname
\fi
#2}}
\providecommand{\BIBdecl}{\relax}
\BIBdecl

\bibitem{richter2022circular}
J.~L. Richter, ``A circular economy approach is needed for electric vehicles,''
  \emph{Nature Electronics}, vol.~5, no.~1, pp. 5--7, 2022.

\bibitem{lin2023economic}
Z.~Lin, P.~Wang, S.~Ren, and D.~Zhao, ``Economic and environmental impacts of
  evs promotion under the 2060 carbon neutrality target—a cge based study in
  shaanxi province of china,'' \emph{Applied Energy}, vol. 332, p. 120501,
  2023.

\bibitem{he2022optimal}
C.~He, J.~Zhu, J.~Lan, S.~Li, W.~Wu, and H.~Zhu, ``Optimal planning of electric
  vehicle battery centralized charging station based on ev load forecasting,''
  \emph{IEEE Transactions on Industry Applications}, vol.~58, no.~5, pp.
  6557--6575, 2022.

\bibitem{zhu2022review}
J.~Zhu, H.~Dong, W.~Zheng, S.~Li, Y.~Huang, and L.~Xi, ``Review and prospect of
  data-driven techniques for load forecasting in integrated energy systems,''
  \emph{Applied Energy}, vol. 321, p. 119269, 2022.

\bibitem{lu2021ultra}
F.~Lu, J.~Lv, Y.~Zhang, H.~Liu, S.~Zheng, Y.~Li, and M.~Hong,
  ``Ultra-short-term prediction of ev aggregator’s demond response
  flexibility using arima, gaussian-arima, lstm and gaussian-lstm,'' in
  \emph{Proceeding of the 3rd International Academic Exchange Conference on
  Science and Technology Innovation (IAECST)}.\hskip 1em plus 0.5em minus
  0.4em\relax Guangzhou, China: IEEE, Feb 2021, pp. 1775--1781.

\bibitem{ren2022hybrid}
F.~Ren, C.~Tian, G.~Zhang, C.~Li, and Y.~Zhai, ``A hybrid method for power
  demand prediction of electric vehicles based on sarima and deep learning with
  integration of periodic features,'' \emph{Energy}, vol. 250, p. 123738, 2022.

\bibitem{zhang2023hybrid}
T.~Zhang, Y.~Huang, H.~Liao, and Y.~Liang, ``A hybrid electric vehicle load
  classification and forecasting approach based on gbdt algorithm and temporal
  convolutional network,'' \emph{Applied Energy}, vol. 351, p. 121768, 2023.

\bibitem{khan2023comparison}
W.~Khan, W.~Somers, S.~Walker, K.~de~Bont, J.~Van~der Velden, and W.~Zeiler,
  ``Comparison of electric vehicle load forecasting across different spatial
  levels with incorporated uncertainty estimation,'' \emph{Energy}, vol. 283,
  p. 129213, 2023.

\bibitem{huang2023explainable}
Y.~Huang, Y.~Zhao, Z.~Wang, X.~Liu, H.~Liu, and Y.~Fu, ``Explainable district
  heat load forecasting with active deep learning,'' \emph{Applied Energy},
  vol. 350, p. 121753, 2023.

\bibitem{benidis2022deep}
K.~Benidis, S.~S. Rangapuram, V.~Flunkert, Y.~Wang, D.~Maddix, C.~Turkmen,
  J.~Gasthaus, M.~Bohlke-Schneider, D.~Salinas, L.~Stella \emph{et~al.}, ``Deep
  learning for time series forecasting: Tutorial and literature survey,''
  \emph{ACM Computing Surveys}, vol.~55, no.~6, pp. 1--36, 2022.

\bibitem{wang2023multivariate}
Z.~Wang, X.~Liu, Y.~Huang, P.~Zhang, and Y.~Fu, ``A multivariate time series
  graph neural network for district heat load forecasting,'' \emph{Energy},
  vol. 278, p. 127911, 2023.

\bibitem{journal/tsg2023/Shi}
J.~Shi, W.~Zhang, Y.~Bao, D.~W. Gao, and Z.~Wang, ``Load forecasting of
  electric vehicle charging stations: Attention based spatiotemporal
  multi-graph convolutional networks,'' \emph{IEEE Transactions on Smart Grid},
  vol. Early Access, 2023.

\bibitem{spiliotis2020cross}
E.~Spiliotis, F.~Petropoulos, N.~Kourentzes, and V.~Assimakopoulos,
  ``Cross-temporal aggregation: Improving the forecast accuracy of hierarchical
  electricity consumption,'' \emph{Applied Energy}, vol. 261, p. 114339, 2020.

\bibitem{journal/access2023/11Mohammad}
F.~Mohammad, D.-K. Kang, M.~A. Ahmed, and Y.-C. Kim, ``Energy demand load
  forecasting for electric vehicle charging stations network based on convlstm
  and biconvlstm architectures,'' \emph{IEEE Access}, vol.~11, pp.
  67\,350--67\,369, 2023.

\bibitem{kaytez2020hybrid}
F.~Kaytez, ``A hybrid approach based on autoregressive integrated moving
  average and least-square support vector machine for long-term forecasting of
  net electricity consumption,'' \emph{Energy}, vol. 197, p. 117200, 2020.

\bibitem{conference/cikm2021/2965cui}
Y.~Cui, J.~Xie, and K.~Zheng, ``Historical inertia: A neglected but powerful
  baseline for long sequence time-series forecasting,'' in \emph{Proceedings of
  the 30th ACM International Conference on Information \& Knowledge
  Management}.\hskip 1em plus 0.5em minus 0.4em\relax Queensland, Australia:
  Association for Computing Machinery, 2021, p. 2965–2969.

\bibitem{dudek2020multilayer}
G.~Dudek, ``Multilayer perceptron for short-term load forecasting: from global
  to local approach,'' \emph{Neural Computing and Applications}, vol.~32,
  no.~8, pp. 3695--3707, 2020.

\bibitem{wang2018short}
Y.~Wang, M.~Liu, Z.~Bao, and S.~Zhang, ``Short-term load forecasting with
  multi-source data using gated recurrent unit neural networks,''
  \emph{Energies}, vol.~11, no.~5, p. 1138, 2018.

\bibitem{8312088}
R.~K. Agrawal, F.~Muchahary, and M.~M. Tripathi, ``Long term load forecasting
  with hourly predictions based on long-short-term-memory networks,'' in
  \emph{Proceedings of the 2018 IEEE Texas Power and Energy Conference}.\hskip
  1em plus 0.5em minus 0.4em\relax College Station, TX, USA: {IEEE}, Feb 2018,
  pp. 1--6.

\bibitem{zeng2023transformers}
A.~Zeng, M.~Chen, L.~Zhang, and Q.~Xu, ``Are transformers effective for time
  series forecasting?'' in \emph{Proceedings of the 37th AAAI conference on
  artificial intelligence}, vol.~37, no.~9.\hskip 1em plus 0.5em minus
  0.4em\relax Washington DC, USA: Association for the Advancement of Artificial
  Intelligence, Feb 2023, pp. 11\,121--11\,128.

\end{thebibliography}
\bibliographystyle{IEEEtran}

\end{document}